\documentclass[usenatbib]{mnras}

\usepackage{newtxtext,newtxmath}

\usepackage[T1]{fontenc}
\usepackage{ae,aecompl}

\usepackage{amssymb,multirow,tabularx}
\usepackage{amsmath}
\usepackage{graphicx}
\usepackage{comment}
\usepackage{BibDef}
\usepackage{xcolor}
\usepackage[normalem]{ulem}
\usepackage{hyperref}
\hypersetup{
	colorlinks=true,        
	linkcolor=blue,         
	citecolor=blue,         
}

\newcommand\msun{M_{\odot}}

\newcommand\LISA{\textit{LISA}}
\newcommand\Zsun{Z_{\odot}}

   \title{Finding binary black holes in the Milky Way with~\LISA}
\author[]{Alberto Sesana$^1$, Astrid Lamberts$^2$ and Antoine Petiteau$^3$\\
$^1$Dipartimento di Fisica ``G. Occhialini'', Universit\`a degli Studi di Milano-Bicocca, Piazza della Scienza 3, 20126 Milano, Italy\\
$^2$Universit\'e C\^ote d'Azur, Observatoire de la C\^ote d'Azur, CNRS, Laboratoire Lagrange, Laboratoire ARTEMIS, France\\
  $^3$APC, Universit\'e Paris Diderot, CNRS/IN2P3, CEA/lrfu, Observatoire de Paris, Sorbonne Paris Cit\'e, 10 rue Alice Domon et L\'eonie Duquet,\\
  \,\,75205 Paris Cedex 13, France\\
}

\date{Accepted XXX. Received YYY; in original form ZZZ}

\pubyear{2019}

\begin{document}
\label{firstpage}
\pagerange{\pageref{firstpage}--\pageref{lastpage}}
\maketitle
 


\begin{abstract}
We determine the main properties of the Galactic binary black hole (BBH) population detectable by~\LISA~and strategies  to distinguish them from the much more numerous white dwarf binaries. We simulate BBH populations based on cosmological simulations of Milky Way-like galaxies and binary evolution models. We then determine their gravitational wave emission as observed by~\LISA~and build mock catalogs. According to our model~\LISA~will detect $\approx4(6)$ binary black holes assuming 4(10) years of operations. Those figures grow to $\approx6(9)$ when models are re-normalized to the inferred LIGO/Virgo merger rates. About 40\%(70\%) of the sources will have a good enough chirp mass measurement to separate them from the much lighter white dwarf and neutron star binaries. Most of the remaining sources should be identifiable by their lack of electromagnetic counterpart within $\approx100$ pc. These results are robust with respect to the current uncertainties of the BBH merger rate as measured by LIGO/Virgo as well as the global mass spectrum of the binaries. We determine there is a 94 per cent chance that~\LISA~finds at least one of these systems, which will allow us to pinpoint the conditions where they were formed and possibly find unique electromagnetic signatures.
\end{abstract}

%

\section{Introduction}
The detection of gravitational waves (GW) from merging binary black holes (BBH) by LIGO/Virgo \citep{LVC_2019_GWTC1} raises the crucial question of the origin of the observed events. 
The first detections reveal a merger rate at  the high end of the theoretical predictions \citep{LIGO:2016_rate,LVC_19_BBHpop_O1O2} and somewhat unexpectedly high BBH masses. These systems likely originate from massive field binary evolution in low-metallicity environments or from N-body interactions in dynamical environments such as star clusters \citep{LIGO:2016_implications}.  Statistical analysis of larger samples of detections may eventually allow to distinguish between these formation channels \citep{Zevin_2017_BBH_channels}. However, given that no electromagnetic counterpart to BBH mergers has been observed so far, the exact identification of the conditions of formation of a given merger remains uncertain.

Stellar mass BHs in the Milky Way have been observed for decades in X-ray binaries. In those systems, the BH is feeding off the companion star, 
and the presence of an accretion disk and of a complementary relativistic jet leads to strong non-thermal emission, from radio to X-rays and sometimes gamma-rays. The masses of the BHs are usually extracted from the Doppler shift in the spectrum of the companion star and found to be between 5 and 10 $\msun$ \citep{Corral_santana_16_BH_catalog}, which differs significantly from the currently observed LIGO/Virgo population. Stellar BHs also cause proper motions to their companion star, and their presence can be inferred by photometric and radial velocity observations \citep{Thompson_19_smallBH,Liu_19_LB1} even if no non-thermal emission is observable 
A handful of unconfirmed BH candidates come from microlensing \citep{Wyrzykowski_16_OGLE_BH}. Finally lone BHs can be lightened-up by accretion of the interstellar medium, although no such objects have been observed so far. As such, our inventory of the BH content of the Milky Way remains very sparse and connecting it to the observed BBH mergers is challenging.


The Laser Interferometer Space Antenna (\LISA) will be a space-base GW detector operating between 10$^{-5}$ Hz and 1Hz. \citet{Sesana16_binaries} showed that certain binaries of masses comparable to GW150914 will be observable by ~\LISA~several months before their merger, and will be mutli-band GW sources.  ~\LISA~will also observe stellar mass compact binaries with periods below one hour within our Milky Way, or nearby galaxies \citep{Korol_18_LISA_LocalGroup}. The vast majority of  these sources will be double white dwarf (DWD), with chirp masses well below 1$\msun$, which will also create an unresolved foreground below a few mHz \citep{nelemans01_WD}. Between 30-300 signals from binary neutron stars (BNS) and a handful of BBHs are likely to be present in the data stream as well ~\citep{Christian17_BBH_LISA,Seto_2019_BNS_MW,Lau_2019_BNS_LISA,Andrews_19_BNS_LISA}. 

Using a binary population synthesis model and a cosmological simulation of a Milky-Way (MW) like galaxy, \citet{Lamberts_18_BBH_MW} showed that roughly a million BBHs are present in the MW. Based on those models, summarized in \S\ref{sec:BBH_pop}, we study in this Letter the population of MW BBHs detectable by~\LISA. We describe the properties of those systems in \S\ref{sec:properties} and strategies to separate them from the outnumbering population of DWDs (and also from BNSs) in  \S\ref{sec:identification}. Finally, in \S\ref{sec:discussion}, we demonstrate the robustness of our results and discuss the scientific payouts of detecting those sources.



\section{The binary black hole population of the MW}\label{sec:BBH_pop}

This work is based on the BBH models presented in \citet{Lamberts_18_BBH_MW}, where all the details about the binary evolution model and the galaxy model can be found. The model is applied to three MW analogs (\textbf{m12i, m12b, m12c}) from the FIRE simulation suite \citep{Hopkins2017fire2}. These models predict that roughly a million BBHs should be currently present in a MW-like galaxy, mostly in the Galactic bulge and stellar halo, as these objects stem from progenitors stars of mean metallicity of 0.25$\Zsun$.

For each of the three galaxies, we generate 100 realisations of the BBH population to assess the statistical uncertainty level in the number of detections due to the stochasticity of the BBH formation process. For each realisation we also randomly choose the phase of the Solar System along its 8 kpc radius circular orbit in the galactic plane. We start our  analysis with 300 BBH catalogs of masses, orbital frequencies and 3D Cartesian coordinates with respect to a reference frame centred on the present-day Galactic center  and with the $x-y$ plane aligned with the galactic disk mid-plane. 

BBHs are assumed to have negligible eccentricity and the sky-inclination-polarization averaged signal-to-noise ratio (SNR) for a BBH of frequency $f$ can be approximated as
\begin{equation}
\bar\rho^2= 2\frac{h^2N}{f\langle S(f)\rangle},
\label{SNR}  
\end{equation}
where $N=f\times T$ is the number of observed wave cycles during the ~\LISA~observing time $T$, $\langle S(f)\rangle =(20/3)S(f)$ is the sky averaged sensitivity of the detector (being $S(f)$ is its intrinsic noise power spectral density), and 
\begin{equation}
h= \sqrt{\frac{32}{5}}\frac{(G{\cal M})^{5/3}}{c^4D}(\pi f)^{2/3} 
\label{hskyave}  
\end{equation}
is the inclination-polarization averaged GW strain. The latter is written as a function of the source chirp mass ${\cal M}=(M_1M_2)^{3/5}/(M_1+M_2)^{1/5}$ and distance $D$. 

We use equation \eqref{SNR} to select the BBHs with $\bar\rho>1$ in each catalog, usually between 30 and 100, depending on the realisation. To each selected system we then assign an inclination angle $\iota$, randomly drawn from a uniform distribution $-1<{\rm cos}\iota<1$, a polarization angle $\psi$ randomly selected between $0$ and $\pi$ and an initial orbital phase $\phi_0$ randomly selected between $0$ and $2\pi$. From ${\cal M}$, $D$ and $f$ we can then evaluate the GW amplitude parameter
\begin{equation}
A= 2\frac{(G{\cal M})^{5/3}}{c^4D}(\pi f)^{2/3} 
\label{amp}  
\end{equation}
and the frequency drift parameter
\begin{equation}
\dot{f}=\frac{96}{5c^5}\pi^{8/3}(G{\cal M})^{5/3}f^{11/3}.
\label{fdot}  
\end{equation}
From the 3D sky localisation we compute the celestial angular coordinates $\theta_s$ and $\phi_s$. Each signal is therefore modeled as a quasi-monochromatic source slowly drifting in frequency and defined by the eight parameters $\vec\lambda=(f,\dot{f},A,\theta_s,\phi_s,\iota,\psi,\phi_0)$. For each system, the signal-to-noise ratio $\rho$ and the uncertainties on each of the parameters $\vec\lambda$ is computed with a code using the Fisher matrix approximation, the core infrastructure of the LISA Data Challenge\footnote{https://lisa-ldc.lal.in2p3.fr/} and a fast computation of the signal in the frequency domain coupling waveform and instrument response. The Fisher matrix approximation has been checked against the bayesian samplers, EMCEE \citep{2013PASP..125..306F} and Dynesty \citep{speagle2019dynesty}. The foreground from unresolved Galactic DWDs is included in the analytic expression of the noise.

\section{Results}
\label{sec:results}
\subsection{Binary Black holes as~\LISA~sources}
\label{sec:properties}
We mark as 'detected' every source with $\rho>7$. Based on our 300 mock populations we expect on average $\bar{N}_4=4.2$ and $\bar{N}_{10}=6.5$ sources to be detected by~\LISA~assuming mission operations of $T=4$\,yr and $T=10$\,yr respectively. The distributions are broad and the median and 90\% confidence intervals of the number of detections are $N_4=3.4_{-3.0}^{+4.9}$ and $N_{10}=5.8_{-4.5}^{+5.7}$. These numbers differ by only $\approx$10\% with those estimated using the more crude sky-inclination-polarization averaged method of Eq.~\eqref{SNR}, which yields $\bar{N}_{10}=5.7$.

\begin{figure}
	\includegraphics[width=.45\textwidth]{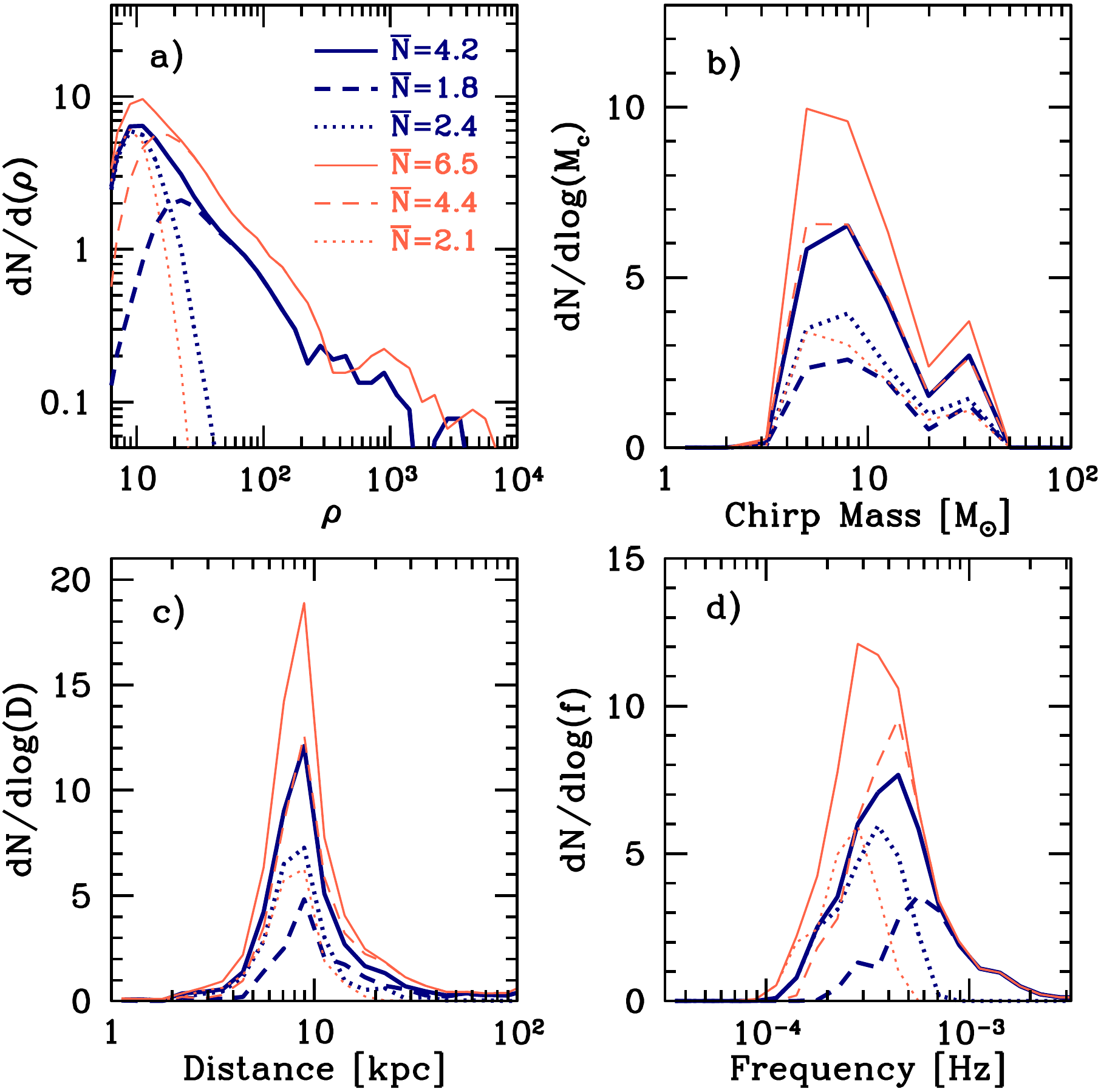}
        \caption{Properties of detected BBHs averaged over 300 galaxy mock realizations: distributions of SNR (a), chirp mass (b), distance to the Sun (c) and  GW frequency (d). Thick--blue(thin--red) lines are for 4(10) years of~\LISA~operations. We plot the total distribution (solid line), and separate systems with (dashed line) and without (dotted line) confident $\cal{M}$ measurement. Panel a) shows the average numbers for each source type.}
    \label{fig1}
\end{figure}

The global properties of the detected sources (signal-to-noise ratio, chirp mass, distance and frequency) are shown in Fig.~\ref{fig1}. Most sources are located 8 kpc away, in the Galactic bulge, and have a frequency of 0.3 mHz, which is $\approx 10$ times lower than the typical frequency at which individual DWDs are detected \citep{nelemans01_WD}. In figure, we highlight sources with confident chirp mass determination, as we will discuss in \S \ref{sec:identification}. 



\begin{figure}
	\includegraphics[width=.5\textwidth]{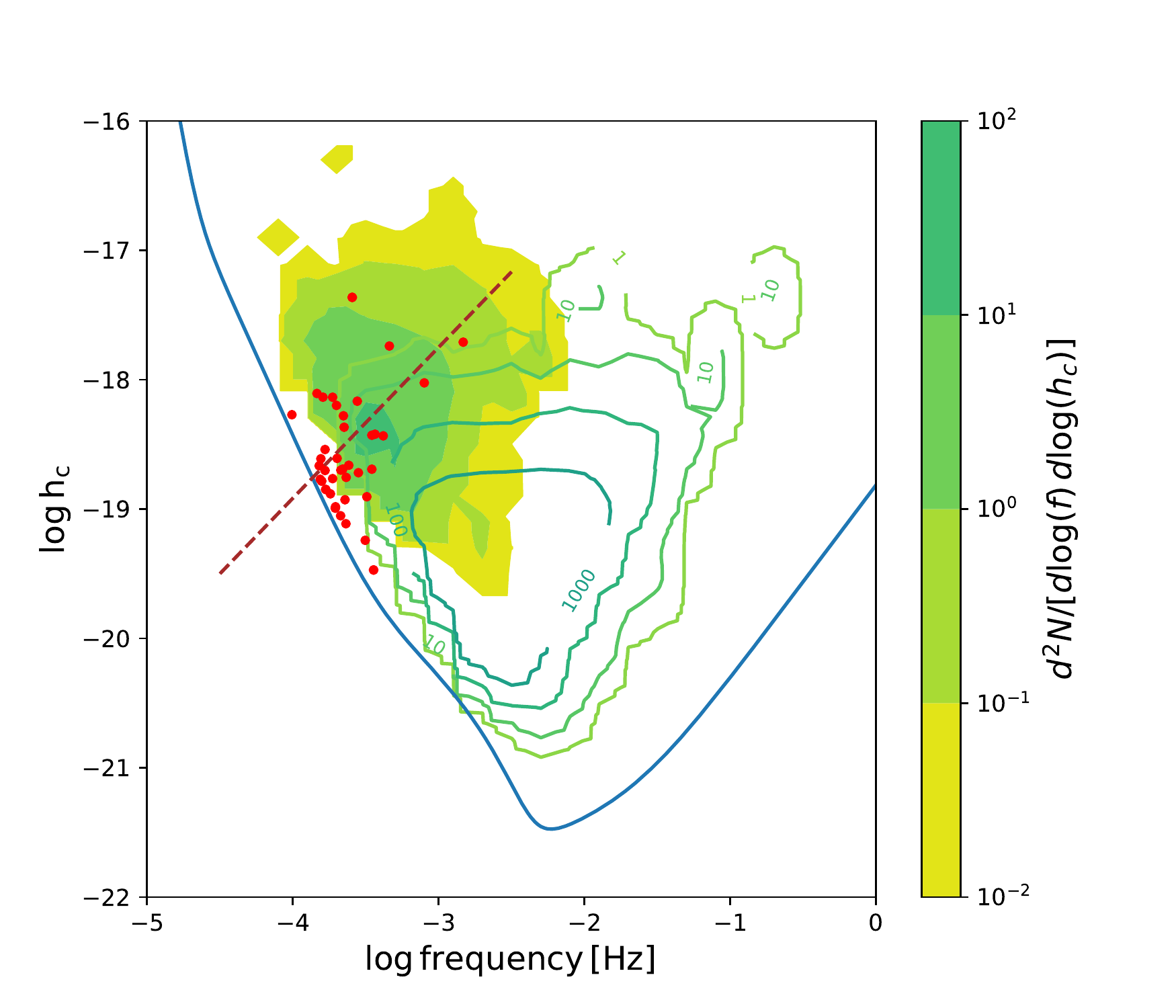}
        \caption{Differential density distribution of BBHs (filled contours, averaged over 300 mock catalogs) and DWDs (open contours) in the $f-h_c$ plane. For~\LISA~we plot the equivalent strain sensitivity $h_{\rm LISA}=f\langle S(f)\rangle$. The red dots display all sources with $\bar\rho>1$ in a selected BBH population for comparison. The brown-dashed line represents the dependence of the charatcteristic strain for a persistent monochromatic source as a function of frequency, $h_c=h\sqrt{N}\propto f^{7/6}$. 10 years of~\LISA~operations are assumed.}
    \label{hc_test}
\end{figure}

Fig.~\ref{hc_test} compares the number density distribution of a mock DWD population with the average number density distribution of observed BBHs from our 300 mock populations in the $f-h_c$ space. $h_c=h\sqrt{N}$ is the charactistic strain of the GW.  The DWD populations is taken from~\citet{Lamberts_19_DWD_MW} and is based on the same Galaxy simulation and binary population model as the BBH distribution. The two distributions are clearly different with the DWDs clustering at 1\,mHz$<f<$10\,mHz, with $10^{-20}<h_c<10^{-19}$, and the BBHs being shifted by an order of magnitude lower in frequency and higher in strain. Nevertheless, the DWD population is much more numerous and it overlaps significantly with the BBH one. All BBHs with $\bar{\rho}>1$ from a selected realization of the Galaxy are also shown for comparison. Although one system on the left can be safely separated from the DWD population, all the others occupy a portion of the parameter space overlapping with the DWD. In the following section, we propose strategies to separate the BBH population from the dominant (number-wise) DWD systems.

\subsection{Identification of GW sources as binary black holes}
\label{sec:identification}
The easiest way to confirm the BBH nature of a detected system is the measurement of its chirp mass. However,  the parameters $\vec\lambda$ of the model do not directly include neither ${\cal M}$ and $D$, which have to be estimated from equations \eqref{amp} and \eqref{fdot}  via error propagation. Assuming for simplicity no correlation among the errors on the parameters, the uncertainties on distance and chirp mass are

\begin{eqnarray}
\frac{\Delta{D}}{D}&=&\sqrt{\left(\frac{2}{3}\frac{\Delta{f}}{f}\right)^2+\left(\frac{5}{3}\frac{\Delta{\cal M}}{\cal M}\right)^2+\left(\frac{\Delta{A}}{A}\right)^2},\\
 \label{deltaD} 
  \frac{\Delta{\cal M}}{\cal M}&=&\sqrt{\left(\frac{11}{5}\frac{\Delta{f}}{f}\right)^2+\left(\frac{3}{5}\frac{\Delta{\dot{f}}}{\dot{f}}\right)^2}.
 \label{deltaM} 
\end{eqnarray}

\begin{figure*}
    \centering
    \begin{tabular}{ccccc}
        \includegraphics[width=0.173\textwidth]{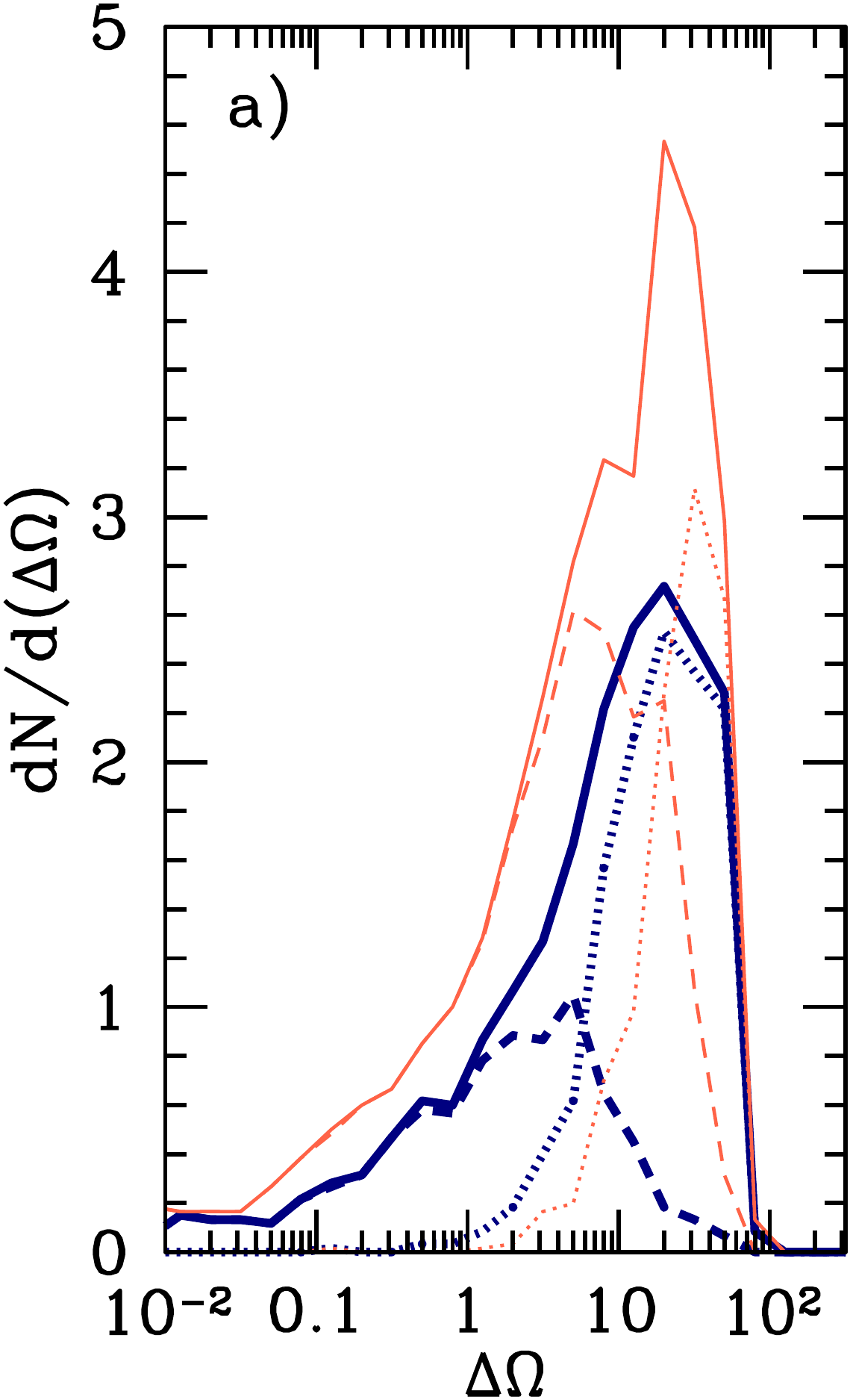}&
        \includegraphics[width=0.18\textwidth]{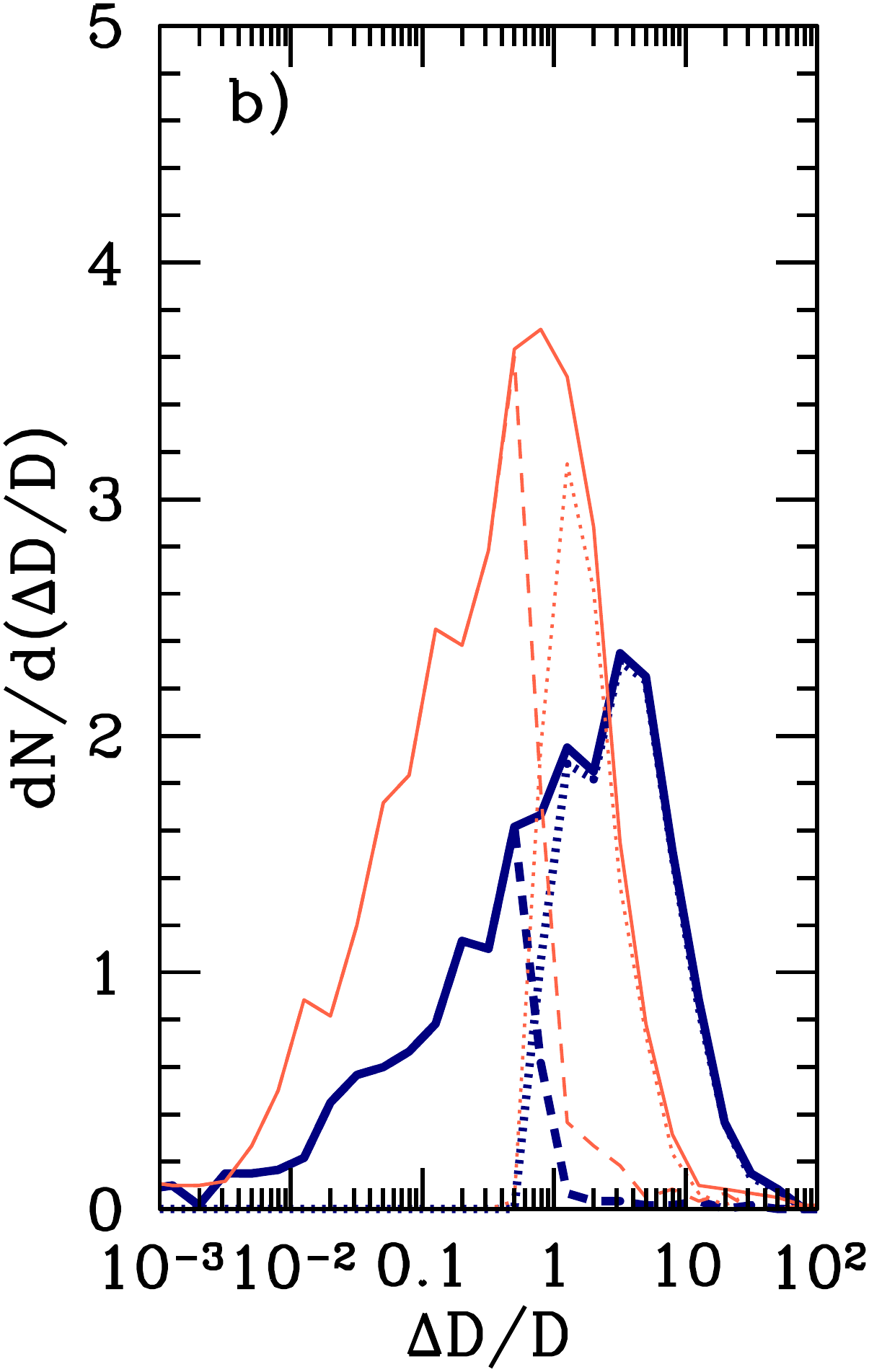}&
        \includegraphics[width=0.18\textwidth]{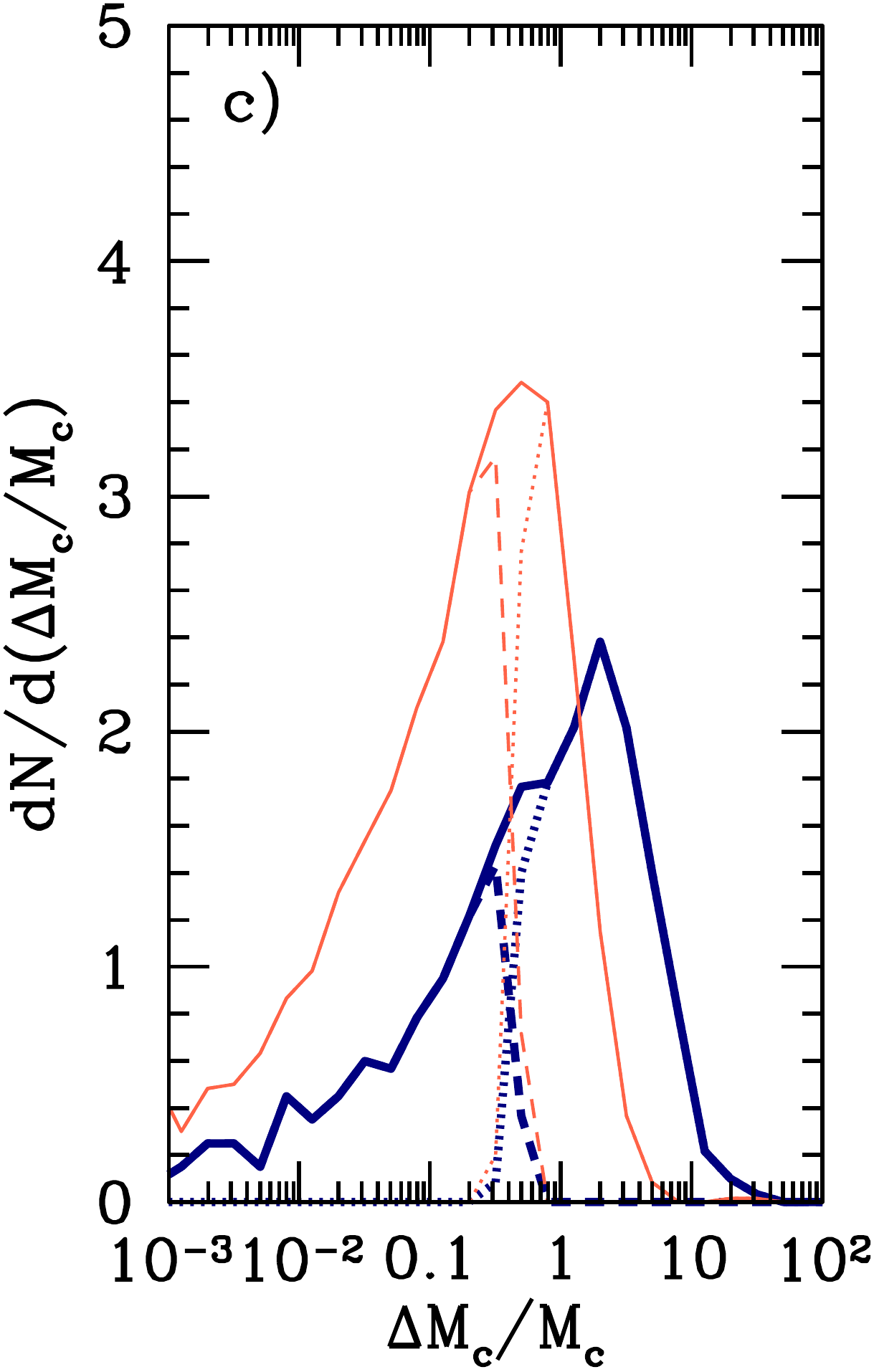}&
        \includegraphics[width=0.18\textwidth]{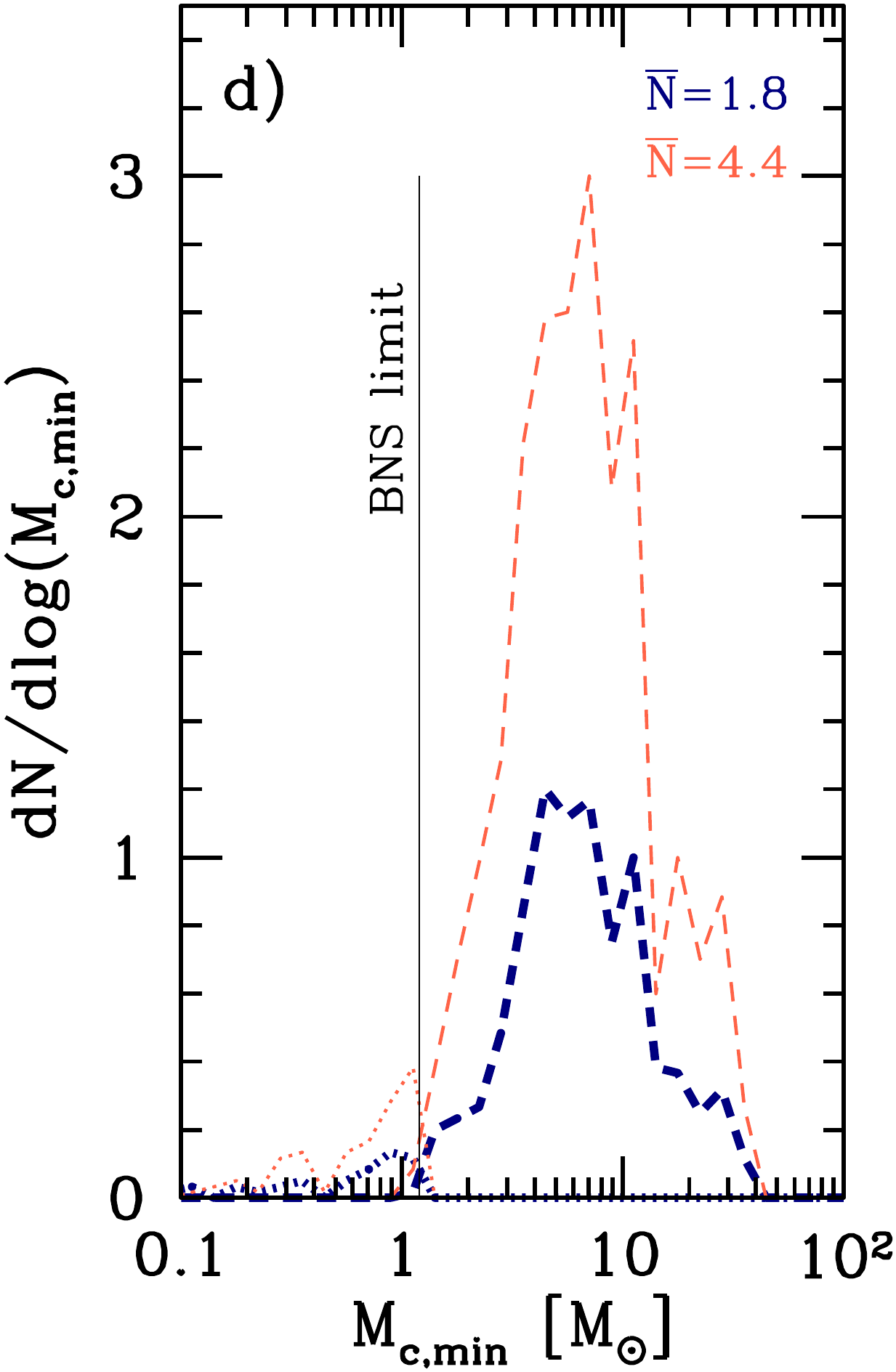}&
        \includegraphics[width=0.17\textwidth]{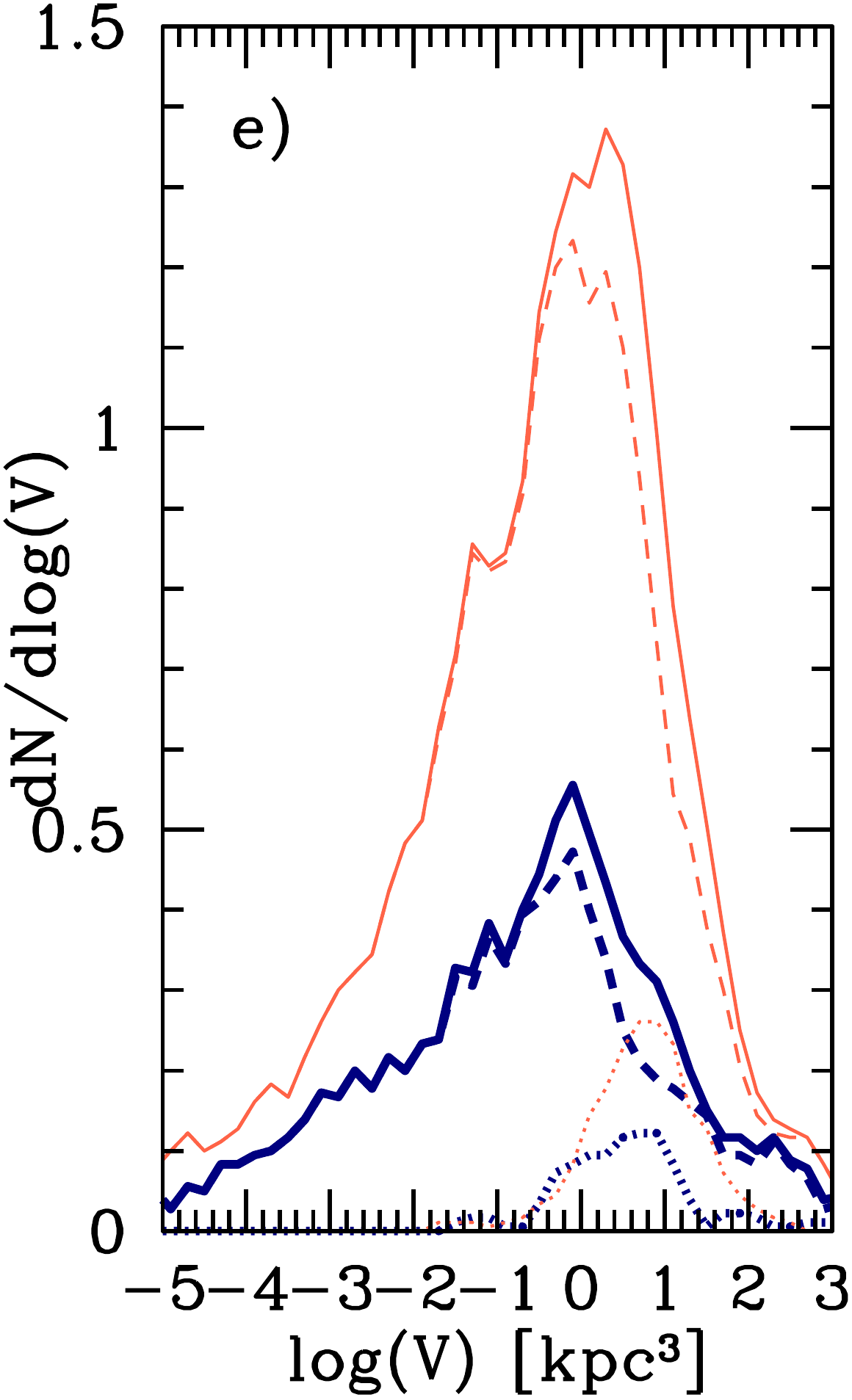}
    \end{tabular}    
    \caption{Measurement uncertainties of the sky localisation (a), distance (b) and chirp mass (c) of the BBHs, averaged over 300 mock catalogs. Panel d) shows ${\cal M}_{\rm min}$ (see text for details): on average, 1.75(4.38) sources have a mass measurement beyond the BNS limit of ${\cal M}=1.2\msun$ for 4(10) year~\LISA~operations. Panel e) shows the 3D volume localization of these well identified BH sources. Linestyle as in Fig. \ref{fig1}.}
    
    

    \label{figparest}
\end{figure*}

Fig.~\ref{figparest} shows the expected measurement error distributions for the sky localization $\Delta\Omega$ (a), distance $\Delta{D}/D$ (b), and chirp mass $\Delta{\cal M}/\cal M$ (c). $\cal{M}$ can be reasonably measured for 40\%(70\%) of the systems for a 4(10) year mission. Using Eq.~\eqref{deltaM} we estimate a 2$\sigma$ lower limit ${\cal M}_{\rm min}={\cal M}-2 \Delta {\cal M}$ which is shown in panel d of Fig. \ref{figparest}.  We identify the sub-population with ${\cal M}_{\rm min}>1.2\msun$, the expected value for a typical BNS. Those are systems containing {\it at least one} black hole and we expect on average 1.8(4.4) of them over a 4(10) year mission.

In all panels of Fig. \ref{fig1} and \ref{figparest} we separate the population of detected BBH with confident mass measurement (${\cal M}_{\rm min}>1.2\msun$) from those without. Fig.~\ref{fig1} shows that the former are generally detected at higher frequency ($f>0.5\,$mHz, panel d) where they have higher $\rho$ (panel a) because of the shape of the~\LISA~ sensitivity curve (cf Fig. \ref{hc_test}). At high frequency, the frequency drift of the signal over the mission lifetime is much larger than~\LISA's frequency resolution, i.e. $\dot{f}\times T\gg 1/T$, meaning that $\Delta\dot{f}/\dot{f}$ in  Eq.~\eqref{deltaM} is small and ${\cal M}$ can be measured with confidence. The direct dependence of this quantity on $T$ explains why ${\cal M}$ can be estimated for a larger percentage of sources in a 10 yr mission. The actual distance or mass of the source have little impact on the measurability of its ${\cal M}$ (see Fig. \ref{fig1}). Fig. \ref{figparest} highlights that systems with measurable ${\cal M}$ are also those with a better estimate of $\Delta\Omega$ and $\Delta{D}/D$ (panels a and b), resulting in a fair estimate of the 3D sky location of the source, shown in panel e). For $T=4(10)$\,yr, we typically expect one(two) sources to be localised within 10 deg$^2$. These sources would also have a 3-D sky localization within a volume smaller than 1\,kpc$^3$.


\begin{figure}
	\includegraphics[width=.45\textwidth]{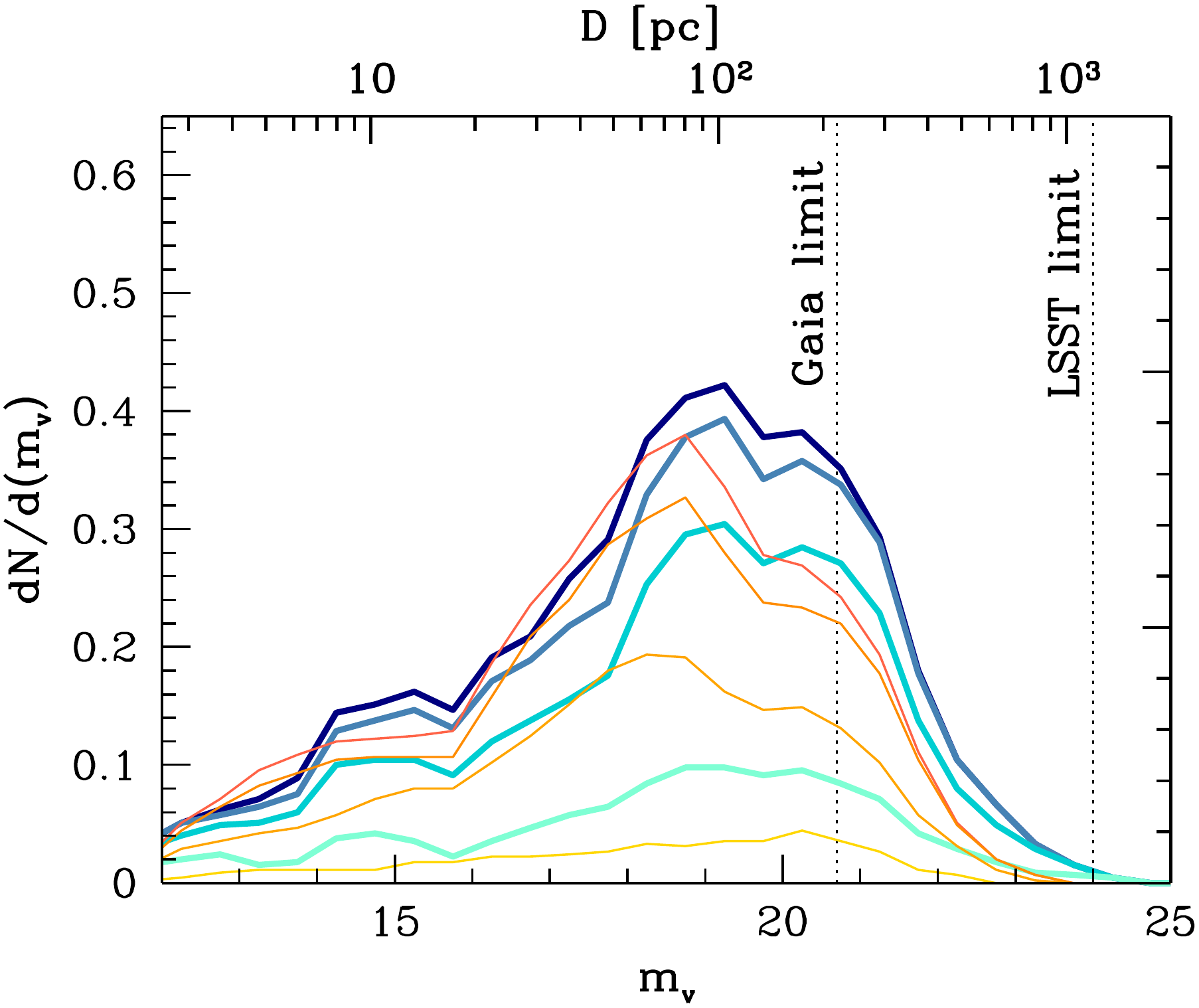}
        \caption{Magnitude (bottom label) and distance (top label) distributions of putative DWDs producing a GW signal equivalent to that generated by the observed BBHs. Only systems at $f<0.5\,$mHz for which ${\cal M}$ cannot be measured are considered. The thick cyan and thin orange lines are for 4 and 10 year of~\LISA~operations respectively. Lines from bottom to top identify systems that can be localized within 10, 30, 50 and 100\,deg$^2$. }
    \label{figmagmax}
\end{figure}

Establishing the BBH nature of sources below 0.5\,mHz is less straightforward and we consider the possibility of identifying them from the lack of an electromagnetic (EM) counterpart. Due to their sub-solar masses, DWDs must be necessarily close to the Solar System to be individually detectable~\LISA~GW sources at  $f<0.5$\,mHz. Assuming a loud GW signal with a given $\hat\rho$ at $f<0.5$\,mHz, for which we cannot measure ${\cal M}$, the maximum distance at which the DWD would lie in order to produce an SNR $\hat\rho$ is   
\begin{equation}
  D_{\rm max}=4\sqrt{\frac{2}{f\langle S(f)\rangle}}\frac{(G{\cal M_{\rm DWD}})^{5/3}}{c^4\hat\rho}(\pi f)^{2/3},
 \label{dmax} 
\end{equation}
where we assume a face-on binary, we use the sky-averaged~\LISA~sensitivity, and we consider a DWD chirp mass of ${\cal M_{\rm DWD}}=0.5\msun$, typical for Carbon-Oxygen WDs which will be the dominant WD population detected by~\LISA~at low frequencies \citep{Lamberts_19_DWD_MW}. Fig.~\ref{figmagmax} shows that DWDs producing contaminant GW signals will be within 600 pc, and mostly within 200 pc. Assuming a conservative absolute magnitude for the WD $M_v=15$ \citep{Gaia_DR2}, the $D_{\rm max}$ distribution can be converted into an apparent magnitude $m_v$ 
\begin{equation}
m_v=M_v+{\rm log}2+5{\rm log}\left(\frac{D_{\rm max}}{10{\rm pc}}\right)
 \label{magmax} 
\end{equation}
where the ${\rm log}2$ factor accounts for the fact that there are two WDs in a binary and we ignore extinction because of the proximity of the sources \citep{Capitanio_17_extinction}. Fig. \ref{figmagmax} shows the apparent magnitude distribution that can be compared to the Gaia single point flux limit of 20.7. The \textit{Gaia} catalog is expected to be complete for WDs up to 60 pc \citep{GentileFusillo_19_WDGaia} beyond which the oldest WDs may not be detected due to their lower temperatures. The WDs detected by~\LISA~have all formed within the last 3 Gyr, otherwise they would already have merged and will thus be detectable at much larger distances \citep{Carrasco_14_GaiaWD}. This effect is the strongest for the faintest most massive Oxygen-Neon WDs which are expected to merge within a few ten million years. Binary interactions such as tidal heating are also expected to increase the flux of such tight binaries. As such, we expect most, if not all of the nearby very short-period DWDs to be present within the Gaia catalog.

The difficulty will be to associate the~\LISA~sources with the WDs from EM the source catalogs. The very short period binaries we consider will not be resolved by Gaia and their binary nature will be unknown. Fig.~\ref{figmagmax} shows that about 50\% of them are located by~\LISA~in the sky with $\Delta\Omega\approx30$deg$^2$, which  corresponds to a volume of $\approx 10^4$pc$^3$ at 100 pc. Given the local WD density (including singles and multiples) of $5\times 10^{-3}$ pc$^{-3}$ \citep{Holberg_16_localWD_density} there will be roughly 50 WDs within the uncertainty region defined by~\LISA. For the least well localised sources, the uncertainty region may contain hundreds of WDs. As such, additional information will be necessary to identify the exact counterpart. Preliminary selection could be done based on the Gaia color-magnitude diagram, ruling out the coldest and oldest stars and possibly subselecting binaries, which are brighter for a given color. Formal identification of a counterpart will require the measurement of a binary period. Such information could be based on the final Gaia lightcurves \citep{Korol17_WDLISA}, follow-up multi-fiber spectroscopic surveys such as WEAVE \citep{Dalton_14_WEAVE}, 4MOST \citep{DeJong_14_4MOST}, or SDSS-V \citep{Kollmeier_17_SDSSV} or lightcurves from high cadence surveys such as ZTF \citep{bellm_2019_ZTF} or  LSST for the faintest systems. 
Given appropriate search strategies, we expect that by the end of the~\LISA~mission, the identification of appropriate EM counterparts to unidentified low-frequency systems will be feasible.

The lack of a plausible DWD candidate would strongly support the BBH nature of the system. DNSs are expected to be rare and unlikely to be detected at $f\lesssim0.3$mHz \citep{Andrews_19_BNS_LISA,Breivik_19_cosmic}. The same is generally true for NS-BH systems which may be more difficult to separate from the BBH population. In any case, we can say with confidence that, in absence of a counterpart, the system contains at least one BH.  


\section{Discussion and conclusions}
\label{sec:discussion}
\begin{figure}
	\includegraphics[width=.45\textwidth]{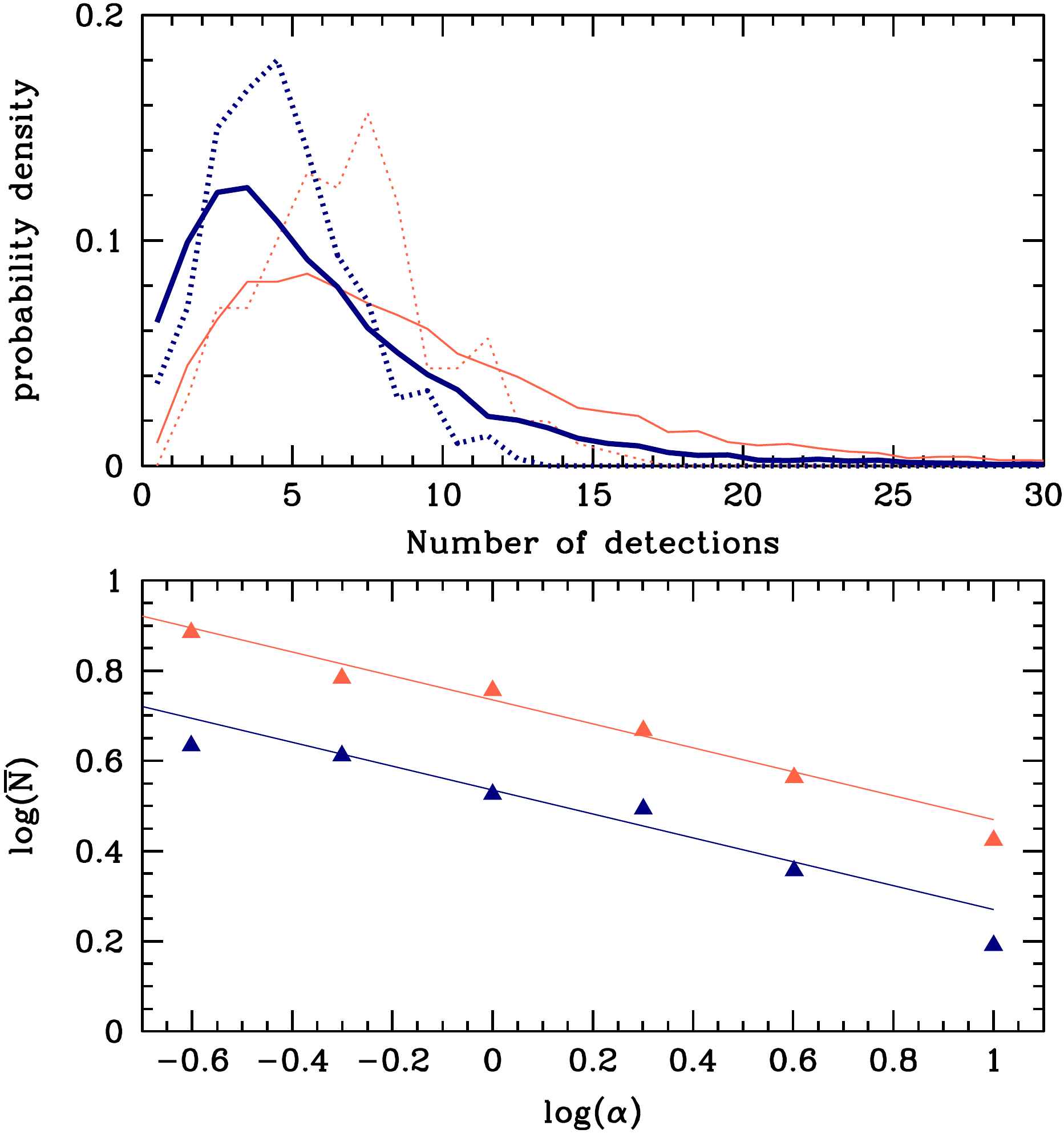}
        \caption{Top: distribution of the expected number of detectable BBHs directly from the 300 mock catalogs (dotted histograms) and after convolution with the LIGO-Virgo measured rate (solid histograms). Bottom: The triangles show the average number of detections from 300 mock catalogs with rescaled BBH chirp masses and the solid lines are the expected analytical scaling (see text). Colorstyle as in Fig.~\ref{fig1}.}
    \label{figdiscussion}
\end{figure}

We base our study on the BBHs found in the three MW equivalent galaxies, which have a BBH merger rate $\approx8\times10^{-6}$yr$^{-1}$. Assuming a MW-equivalent volume density of 0.005 Mpc$^{-3}$ \citep{Tomczak:2014}, this results in a BBH merger rate at $z=0$ of 40 yr$^{-1}$ Gpc$^{-3}$, which is consistent with the measured LIGO-Virgo rate of $53^{+58}_{-29}$yr$^{-1}$ Gpc$^{-3}$ \citep{LVC_19_BBHpop_O1O2}. To fold into our calculation the uncertanties in the measured BBH rate, we convolve the distribution of expected detections with the  posterior distribution of the LIGO/Virgo merger rate. Fig. \ref{figdiscussion} shows that based on this, we expect an average number of detections of $\bar{\cal N}_4=6.1$ and $\bar{\cal N}_{10}=9.3$. The distributions are highly asymmetric and the median and 90\% confidence intervals of the number of detection are ${\cal N}_4=4.8_{-4.0}^{+10.7}$ and ${\cal N}_{10}=7.7_{-5.8}^{+14.8}$. The probability of \LISA~detecting {\it at least} one BBH is 0.94 and 0.99 in the two cases. This prediction holds under the assumption that the dominant BBH formation channel is binary field evolution. A significant contribution from a range of dynamical channels might in fact produce very eccentric binaries, resulting in sparser sources in the~\LISA~band \citep[e.g.][]{2017MNRAS.465.4375N}. 

We investigate how our results depend on the model of the BBH chirp mass distribution. Naively, a factor of two difference in ${\cal M}$ would result in a factor $>3$ increase in the maximum distance of observable sources and thus in a factor $\approx 30$ difference in the number of detected systems. This is in general not the case, as we shall now demonstrate. In all cases, the merger rate has to satisfy the LIGO constraint. Since $dn/d{\rm ln}f=(dn/dt)(dt/d{\rm ln}f)$, at a fixed rate, the number of binaries per unit log frequency is proportional to the time they spend at that frequency, i.e., $dt/d{\rm ln f}\propto f^{-8/3}{\cal M}^{-5/3}$ (cf Eq. \eqref{fdot}). It results that the number of detectable sources is set by the lowest observable frequency $f_{\mathrm{min}}$. Let us consider a source at a given distance; its characteristic strain integrated over the observation time is $h_c=h\sqrt{N}\propto f^{7/6}{\cal M}^{5/3}$ -- cf Eq. \eqref{hskyave}) --, as represented by the dashed-brown line in Fig. \ref{hc_test}. The line intersects the sensitivity curve at a frquency $f_{\mathrm{min}}$, that depends on the considered source and chirp mass.
Since $h_c\propto f^{7/6}{\cal M}^{5/3}$ and the~\LISA~sensitivity in the relevant frequency range is $h_{\rm LISA}\propto f^{-2}$, by taking the log of the two expression and equating them, one gets ${\rm log}f_{\mathrm{min}}\propto-(10/19){\rm log}{\cal M}$. Since ${\rm log}N\propto-(8/3){\rm log}{f}-(5/3){\rm log}{\cal M}$, this gives  ${\rm log}N\propto-(15/57){\rm log}{\cal M}$. This means that a change in chirp mass by a factor of 2  only changes the number of detected events by about 20\%. To test this, we artificially multiply the chirp mass of all the BBHs in our catalog by a factor $\alpha$ (re-weighting them by a factor $\alpha^{-5/3}$ in order to preserve the merger rate) and compute the number of ~\LISA~ detections. Even when we vary $\alpha$ by almost two orders of magnitude, the number of detected sources is close to our current model for both 4 and 10 yrs LISA mission, as shown in the lower panel of Fig. \ref{figdiscussion}.
We therefore conclude that the number of detections estimated here are robust and only mildly dependent on the detailed properties of the BBH mass distribution.  

The detection of Galactic BBHs therefore sets another important goal of the~\LISA~mission. The determination of their chirp mass and 3D localization within the MW might provide important clues about their origin, and their connection to other galactic BHs found in X-ray binaries. With this goal in mind, we stress the importance of an extended~\LISA~lifetime. When normalized to the LIGO/Virgo merger rate 10 years of~\LISA~operations will allow the detection of ${\cal O}(10)$ binaries, and relevant parameter measurements for the majority of them. Finally, with ${\cal O}(1)$ system localized within 1 deg$^2$, the~\LISA~Galactic BBH detections may also offer the first opportunity to observe any EM counterpart to isolated BHs, including radio observations with SKA or X-ray observations with the \textit{Wide Field Imager }on-board of \textit{Athena} satellite.

\section*{Acknowledgements}   
This research was supported in part by the National Science Foundation under Grant No. NSF PHY-1748958. AS is supported by the ERC through a CoG grant. AL acknowledges support by the Programme National des Hautes Energies (France). AP acknowledges support by the Centre National d'Etudes Spatiales. The authors acknowledge the LISA Data Challenge Team that developed the core part of the code used for parameter estimations.
\bibliographystyle{mnras}
\bibliography{sources}
%
%

\end{document}